\documentclass{llncs}
\usepackage{graphicx}
\usepackage{listings}
\usepackage{hyperref}

\begin{document}
\title{Reverse Engineering of RFID devices}
\author{Wouter Bokslag}
\institute{Department of Information Security Technology\\Technische Universiteit Eindhoven}
\maketitle

\begin{abstract}
This paper discusses the relevance and potential impact of both RFID and reverse engineering of RFID technology, followed by a discussion of common protocols and internals of RFID technology. The focus of the paper is on providing an overview of the different approaches to reverse engineering RFID technology and possible countermeasures that could limit the potential of such reverse engineering attempts. 
\end{abstract}

\section{Introduction}
\label{sec:introduction}
The definition of reverse engineering is, according to the Merriam-Webster dictionary, "to disassemble and examine or analyse in detail (as a product or device) to discover the concepts involved in manufacture usually in order to produce something similar" \cite{web:merweb-def}. Although the exact term has only recently been adopted, the practice of discovering the properties and internals of objects manufactured by another party has been practised for ages. Many early inventions have been copied by other cultures, as they brought some specific benefit to those who mastered the technology\cite{web:history}. Often, these advantages were of a military nature, as most civilizations or nations strive to have a strategic edge over their rivals. Nowadays, there are a multitude of motivations for reverse-engineering. The principal ones will be discussed briefly below.

\subsection{Espionage and/or cloning}
Espionage is one of the first reasons that come to mind when discussing the purpose of reverse engineering. Both enterprises and nations are interested in the technologies and products other parties have developed. This interest has multiple reasons, the first of which being able to assess the full capabilities of the product. Another common reason is to be able to either copy the entire product or use a subset of the design features for use in other or similar products. The third reason is mainly relevant for corporations: deriving the production cost. If a corporation has knowledge of the manufacturing cost of their competitor's products, they are able to formulate a competing strategy. For instance, when it is known that a competitor has only little margin on a product, it might be interesting to decrease the price of a similar product in order to either reduce the competitor's sales or force them into selling the product for (or even under) the cost price\cite{doc:stateoftheartIC}. Also, it might allow a party to derive how to produce a product or parts of that product, allowing the reverse engineer to avoid the need for research and development. 

\subsection{Analysis of legacy or poorly documented products}
Nowadays, there are large amounts of software that are no longer actively maintained by the original developers. When documentation is available or the software is open-source with annotated, high quality code, this might not be a problem. Often however, this is not the case, and reverse engineering may be employed to shed light on both the quality and the inner workings of software. This way, documentation may be compiled for previously unmaintainable software, allowing the continued use and/or development of the software. A similar motivation exists for physical products: finding out details about the quality of the assembly, materials used, etcetera\cite{web:altertech}. 

\subsection{Circumventing copyright infringement}
While international copyright treaties consider copying a software product as copyright infringement, there exists a reverse engineering-centred approach that can circumvent this restriction. This approach, called clean room design, requires the work two completely separate teams. The first team 
 the software and writes a complete specification of its operation. The second team creates a software product based solely on the specification written by the first team. The term 'clean room design' refers to the fact that the end product can be shown to be free of any 'contamination': the implementing team had no knowledge of the structure and design of the original software product. One notable example is the reverse engineering of the IBM BIOS by Phoenix Technologies\cite{ebook:sipp}, allowing Phoenix to build IBM compatible computers (and license its BIOS to other parties as well). 

\subsection{Verification of security}
In many cases, the security claims a manufacturer makes about a closed-source software or hardware product must be taken at face value. While black box penetration tests can provide an indication of the security of the product, reverse engineering approaches that uncover and assess the internal structure of the product may reveal vulnerabilities that would otherwise have remained undiscovered. The reason is that black-box penetration testing is limited to the regular input-output channels, and although fuzzing may be an effective way to discover some potential security issues, it can by no means detect \emph{all} (classes of) issues. Intentional backdoors that are added by the manufacturer often rely on some secret (combination of) parameter(s) that trigger a high-privilege mode, but also non-documented debug-functionality may be present\cite{doc:jtag}. Reverse-engineering may uncover such 'functionality', as the product will somewhere implement this functionality. In a software product, there must be code that checks for the secret and enters a high-privilege mode, while a hardware product will have a set of registers, gates and circuits that cannot be attributed to any of the 'public' functionality.


\section{On RFID}
\label{sec:rfid}
\begin{figure}[ht!]
\centering
\includegraphics[width=130mm]{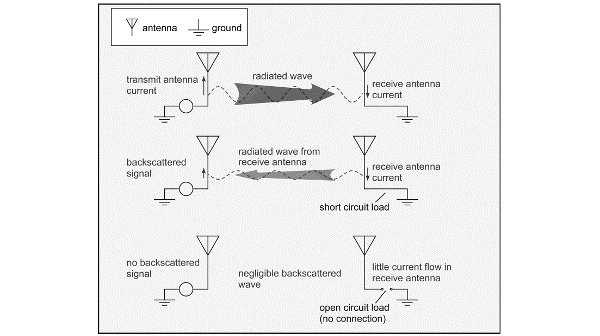}
\caption{The reader sends the binary string 10100001000. The field reflected by the tag depends on the bits sent by the reader just before. Source: \cite{web:embedded}}
\label{fig:backscatter}
\end{figure}

RFID is a technology that allows for communication between a tag and a reader. This paper will assume the reader to be active and the tag to be passive. This is by far the most common setup, in which the reader creates an alternating electromagnetic field that is used as a power source by the tag. The electromagnetic field induces a voltage on the tag's antenna, and by alternating the draw of current from the antenna the \emph{backscatter} can be controlled by the tag, allowing the reader to receive information from the tag\cite{web:eetimes} as shown in illustration \ref{fig:backscatter}.

RFID can operate on a range of frequencies. The most common ones range from 120KHz (LF) up to 10GHz (microwave), where the higher frequencies can obtain higher data transfer rates, but at the cost of a reduced range. In this paper we will focus on smartcards, which operate around 13.56MHz and follow ISO/IEC 14443\cite{doc:iso14443}. This standard specifies the physical characteristics, radio power and frequency, anti-collision and the transmission protocol that have to be used. While many RFID tags only provide a single read-only serial number, many RFID smartcards have read/write support as well as crypto and authentication functionality. 

\section{Approaches to reverse-engineering RFID}
\label{sec:approaches}
In this section, different approaches to reverse engineering that are relevant for RFID are discussed. A distinction is made between non-intrusive and intrusive approaches. The intrusive methods require physical access to the internals of the RFID chip and, therefore, will inevitably destroy the tag. 

\subsection{Non-intrusive approaches}
There are several non-intrusive approaches to reverse engineering. The most important ones are protocol analysis, and EM/power analysis. While usually two distinct classes of attacks, in the context of RFID EM and power analysis cannot be seen separately.

\subsubsection{Protocol analysis}
One of the most popular approaches to reverse engineering is protocol analysis. The information used for discovering the internals of the tag is limited to communication traces between the tag and a reader. 

The popularity of protocol analysis is easily explained. First, there is no or little additional hardware required, making this one of the cheapest approaches. Also, if one is able to derive the protocol used for communication between reader and tag, one has effectively derived the most important part of the functionality. 

In some cases the attacker is unable to directly control the contents of the communication towards the tag, as the reader may contain a proprietary controller that implements part of the communication protocol. However, even if a proprietary, closed-source controller is being used, analysing the communication between tag and reader is still a viable approach to uncovering the protocol\cite{web:snifferonly}. 

It is often very hard to derive \emph{all} of the functionality, as it is difficult to reverse-engineer, for instance, cryptographic algorithms. Often, the tag has registers containing secret values that are used for challenge-response and cryptographic algorithms. Due to the nature of such algorithms the secret values cannot be recovered through knowledge of the algorithm and the output, but it seems clear that recovery of the algorithm with only knowledge of the key is equally difficult. However, even partial information about the communication protocol can be of interest, as it sheds light on the design of the tag, information that could be helpful when attempting other approaches of reverse engineering. 

\subsubsection{(Power / EM analysis}
\label{sec:approaches-power}
While generally two separate classes of side channel attacks, within the scope of passive RFID tags, power and electromagnetic radiation based attacks are indistinguishable. This is because the power source of the tag \emph{is} the electromagnetic field generated by the reader, and its power consumption can only be estimated by analysing the electromagnetic field caused by the tag's antenna. Although thus different from 'traditional' power and EM analysis, this still is an interesting approach to attacking and reverse-engineering RFID tags. Yossi Oren and Adi Shamir were the first to show that power analysis is indeed a viable approach to attacking tags\cite{doc:yossen-shamir_poweranalysis}. 

One of the main concepts of power analysis is that the power consumption of a hardware device is usually roughly proportional to the amount of bits that change their value at a certain time. This allows to distinguish when the device is idle and when it is busy, helping to find out what the device is doing at a certain time. 

\begin{figure}[ht!]
\centering
\includegraphics[width=130mm]{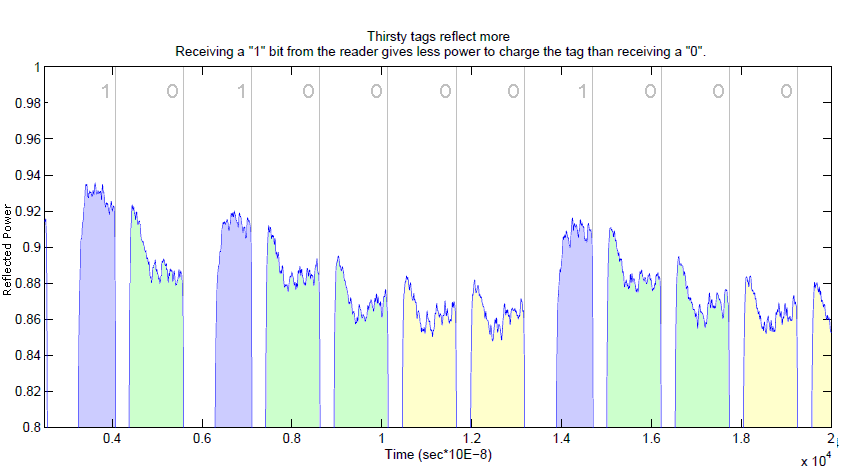}
\caption{Relation between data sent to tag and reflected power. Source: \cite{doc:yossen-shamir_poweranalysis}}
\label{fig:thirstytag}
\end{figure}

To understand how power analysis of RFID works, it is important to understand the relation between the electromagnetic field caused by the reader, and the way the tag can influence this field. The strong field generated by the reader causes an electrical current to flow back and forth in the tag's antenna. This alternating current is amplified, rectified and stored in the card's internal power storage. In this way, the card has direct-current power available for the main circuitry. Because a current is flowing through the antenna, the tag also generates an electromagnetic field. The strength of this field is governed by the current flowing through the antenna, which equals the power consumption of the tag. For clarity, the measured 'power consumption' is thus equal to the amount of power that is used to charge the internal power storage. Maximum charge generates a powerful electromagnetic field, while low charge only generates a weak field. When the reader is transmitting information to the tag, it does so by alternating a 'high' field with no field. The duration of the high field determines whether a 0 or a 1 is being transmitted: a long energized field is defined as a 0, while a short field (and thus preceded by a relatively long gap) means a 1, as illustrated in figure \ref{fig:thirstytag}. 

Note that when the reader transmits a 0, there is a longer gap and a relatively short pulse. This means there is less power available for the tag when receiving a 0, resulting in a larger drain from the internal power storage. The important conclusion to draw from figure \ref{fig:thirstytag} is that a tag that has depleted a large part of its power reserves is more 'thirsty': it draws more power from the next pulse it receives, generating a stronger electromagnetic field. 

\begin{figure}[ht!]
\centering
\includegraphics[width=130mm]{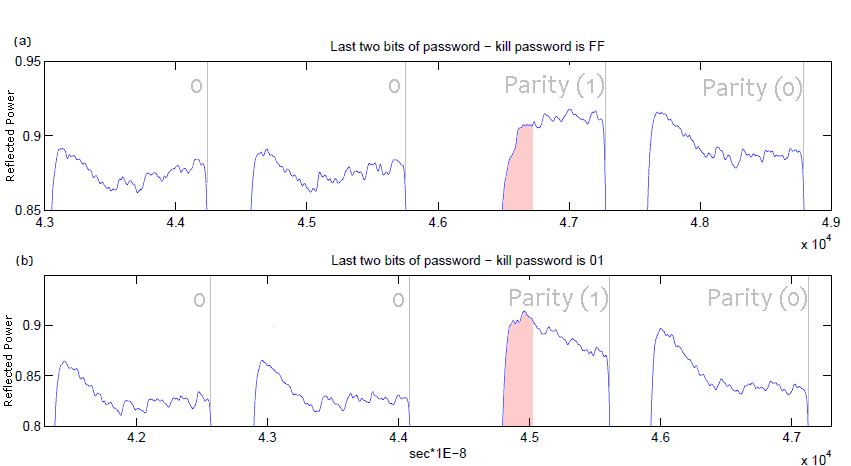}
\caption{The reader sends the kill-command with key 0000 0000 and parity bit 1. The first card expects the key 1111 1111 while the second card expects 0000 0001. Source: \cite{doc:yossen-shamir_poweranalysis}}
\label{fig:thirstyPOC}
\end{figure}

As a proof of concept of a power analysis that exploits this different behaviour, Yossen Oren and Adi Shamir\cite{doc:yossen-shamir_poweranalysis} sent a kill-command with password 0000 0000 and parity bit 1 to two RFID-cards. The first card expected a very different password: 1111 1111. The second card, on the other hand, expected the password to be 0000 0001, implying that this card can only detect that the supplied password was incorrect when comparing the very last bit of the password. This suggests that the second card will require more power when processing the last bit of the password, which should be visible in the strength of the reflected field from the parity bit: a thirsty card emits a stronger field. Figure \ref{fig:thirstyPOC} shows that this is indeed the case. This clearly shows that power analysis is a viable approach to discovering information about the internal state of RFID tags. 

The above approach was developed to work on UHF tags, which operate in the 900MHz frequency range. However, similar attacks are also possible on 13.56MHz RFID devices. Michael Hutter et al. have already been able to perform successful differential power attacks against different RFID tags, allowing them to recover AES encryption keys\cite{doc:hutter_powerEManalysis}. 

\subsection{Intrusive approaches}
Two distinct intrusive approaches to reverse-engineering will be discussed. The first one is optical analysis, which is used widely to reverse-engineer microchips. The second approach that will be considered is electronic analysis, where the electrical signals inside the chip are probed. 

\subsubsection{Optical analysis}
Optical analysis is an approach where the interior of the chip is being analysed by means of imaging. Different technologies are shared under this name. In this section, the most commonly used ones are discussed briefly. Because access to the silicon is required, the chip package has to be removed. This is generally done using chemicals like fuming acid or the less aggressive acetone. For analysis of all layers, fine grain polishing solutions may be used to grind of tiny layers at a time. However, due to the physical stress this induces on the chip, more advanced solutions employing lasers or focused ion beam generally yield better results. 

\begin{figure}[ht!]
\centering
\includegraphics[width=130mm]{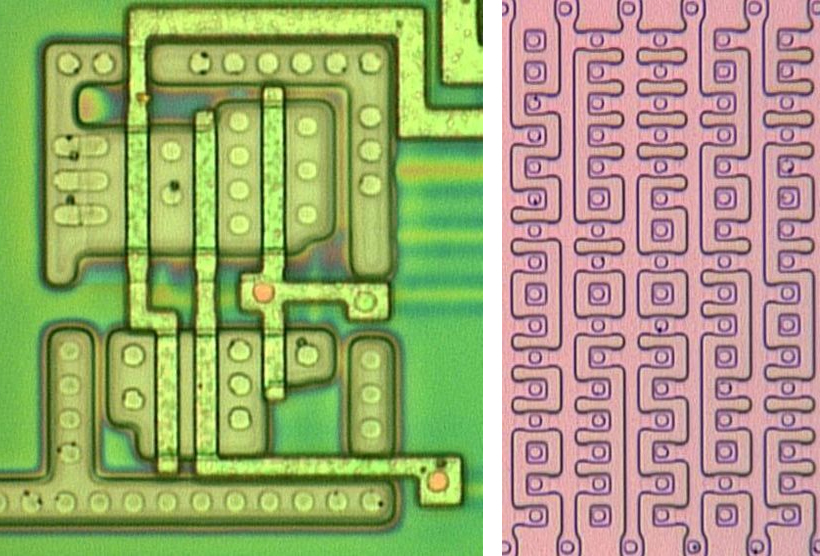}
\caption{Left: AND-gate, after removal of metal layer by etching. Right: 16x10 bit ROM. The value of each bit is easily visible. Source: \cite{doc:kommerling}}
\label{fig:gate_rom}
\end{figure}

First, there is the optical microscope. It must be noted that the smallest visible detail of optical microscopes is governed by the wavelength of the light that is being used to illuminate the sample. The smallest visible detail on optical microscopes using conventional lenses is around 200nm. Where 22nm technology is already being used and technology for 14nm processes are already suitable for mass production\cite{web:14nm}, the limitations of optical microscopes are obvious. However, many RFID tags are made with older generations of lithographic equipment, and optical microscopes may still be very useful when reverse-engineering RFID tags. As illustration \ref{fig:gate_rom} shows, details can be very clearly visible. The right part is a 16x10 bit ROM circuit, the contents of which are easily identified. Each bit is a connection from one of the 6 vertical lines to one of the 5x16 dots, that connect to the diffusion layer. If no such connection exists for a given line/dot combination, the value of that bit is 0. 

\begin{figure}[ht!]
\centering
\includegraphics[width=130mm]{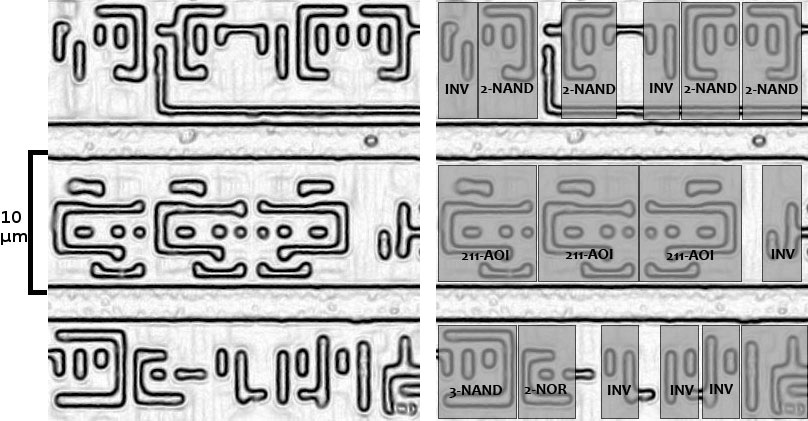}
\caption{A detail of the MiFare Classic RFID tag. Left: the image after edge detection. Right: Image after automatic template detection has been performed. Source: \cite{doc:karstennohl}}
\label{fig:mifareclassic}
\end{figure}

Karsten Nohl et al.\cite{doc:karstennohl} used a non-modified standard optical microscope with a 500x magnification factor when reverse-engineering the MiFare Classic RFID tag, grinding away layer by layer the surface of the RFID tag with polishing emulsion and very fine sandpaper. The fact that the different layers of the chip are very close together proved to be a problem, as they could not manage to completely eliminate any tilt of the chip. This means that grinding will always cut through multiple layers on the chip, resulting in the impossibility to photograph a layer completely on one picture. Using modified panorama stitching software, they managed to overcome this complication without the need for advanced stabilizing and anti-tilt equipment. Using automated pattern recognition and a templating engine, illustrated in figure \ref{fig:mifareclassic}, they proceeded to identify the different gates on the chip and started looking for the crypto logic. 

They knew \emph{some} details about the algorithm, for instance, that a 56-bit key was used. For this reason, they started to look for a 56-bit register, which they found alongside what they identified to be a random number generator. Inspection learned them that the RNG had no input circuits, which implied that no proper seeding was being done. Indeed, it later became clear that the RNG only depended on the time since startup, which is highly predictable and proved to be a crucial design flaw. 

More commonly used by professional reverse engineering companies are different variants of electron microscopes. Scanning electron microscopy can provide a highly detailed view of the surface of a chip, while a transmission electron microscope can be used on very thin 'slices' of a sample. Another very interesting technology is scanning capacitance microscopy that, just like energy-dispersive X-ray spectroscopy, can give information about the materials and alloys used.

\subsubsection{Electronic analysis}
While optical analysis can reveal in great detail the layout of the circuitry of a tag, it cannot always provide sufficient information to comprehend its workings. Electronic analysis consists of placing tiny probes on the chip, that allow to read the value of a circuit during operation. While relatively simple on single or dual-layer chips, modern multi-layer chips are harder to analyse as large parts of the chip are not accessible. In these cases, a focused ion beam may be used to remove chip material an 'dig a hole' in non-crucial parts of the chip surface. This way, probes can be attached even to parts of the chip that are normally covered by superposed layers. 

Advanced, carefully crafted tricks such as intentional damage to specific circuits, voltage fluctuations and variable clock speeds can trick the chip to \emph{glitch}. This can, for instance, change the branch of a conditional jump, allowing the attacker to bypass built-in protection mechanisms.


\section{Countermeasures}
\label{sec:countermeasures}
It has been shown that reverse engineering can be employed to discover the structure and inner workings of RFID tags. Personally, I firmly believe that transparency and open-source design will lead to faster innovation and will allow researchers to find and publish vulnerabilities in an earlier stage. This is an advantage, as it allows many potential customers to consider the security risks \emph{before} adopting a particular RFID solution, decreasing the risk that major flaws are discovered when the whole system is already deployed. However, it cannot be denied that in some cases, transparent, peer-reviewed and open-source design is not an option. Even manufacturers that decide to rely on security through obscurity will be interested to implement countermeasures to common reverse-engineering approaches. It must be noted, however, that these countermeasures are often costly, as they require additional transistors or other hardware equipment, or simply complicate the design process of the tag. In this section, countermeasures are presented against each of the attacks discussed in section \ref{sec:approaches}. 

Please note that this is by no means an exhaustive list of countermeasures. Many, many solutions have been presented, but it would be outside the scope of this paper to consider all alternatives. A good general overview to the subject is presented in \cite{doc:kommerling} by Oliver Kömmerling.

\subsection{Countermeasures against protocol analysis}
Although it is impossible to fully protect against protocol analysis, it can be severely hindered. We consider the case where the RFID reader implements the proprietary protocol, for when open or known protocols are used, an attacker has no need for protocol analysis in the first place. One important approach to keeping the protocol secret is by implementing strong crypto primitives that obfuscate the communication as soon as possible. After an initial handshake, authentication and/or key-exchange, the communication can be fully encrypted, requiring the attacker to obtain secret keys and/or information about the used algorithms in another way. It must be noted that although relatively simple, the solution requires additional logic. Secure cryptographic algorithms must be implemented, as well as registers that can hold keys and intermediate values. This increases the cost of the RFID tag and might also increase power requirements (which might decrease the operational range) and have a negative impact on both latencies and transfer speeds.

\subsection{Countermeasures against power/EM analysis}
Preventing basic power analysis can be done by implementing an important rule of thumb: for every branch that must remain secret to the attacker, both branches should result in the same amount of \emph{work}. Consider the proof-of-concept by Yossen Oren and Adi Shamir discussed in section \ref{sec:approaches-power}. Power analysis worked because the internal logic stops comparing bits as soon as it finds that the password is not correct. What \emph{should} be done in order to prevent power analysis is comparing each single bit, while checking afterwards if one or more bits were incorrect. In order to prevent the attacker for discovering when the 'wrong key'-bit was set, one could choose to implement two counters, counting the number of correct and wrong bits respectively. This way, each bit, wrong or correct, leads to exactly the same amount of work and thus results in identical power usage. Pseudo-code of this approach is provided below. 
\pagebreak
\begin{lstlisting}[frame=single]
// Key will be compared with reader-supplied 'test'
// Note that key[4] denotes bit in variable key
key =  11111111
test = 00000000

correctCount = 0
wrongCount = 0

for i = 0; i < 8, i++
begin
	if (key[i] == test[i]) then
		correctCount++
	else
		wrongCount++
	end
end

if (wrongCount == 0) then
	echo "success!"
end
\end{lstlisting}

It must be noted that this is an expensive solution in terms of circuitry, as additional logic and registers must be implemented that serve no function other than assuring a constant energy usage. Other solutions are possible, such as circuits that have a random power consumption, but it is very difficult to get this solution to remain effective when multiple traces are available to the attacker. 

\subsection{Countermeasures against optical analysis}
It is not easy to protect against optical analysis, as different powerful imaging technologies exist. Randomizing the layout of gates is a good idea, for this makes the automated analysis of the functionality of the chip harder. Also, decoy circuits (circuits that seem to be functional, but actually consist of 'dead' gates that do not add any functionality could be added. Another approach is to store key parts of the functionality in memory (that is, semiconductor-based memory cells, as these cannot be read optically), as optical methods cannot determine the contents of registers and memory cells. Either crucial values or entire algorithms can be hidden from optical analysis. The cost of all of these approaches, however, is significant in terms of development effort and number of transistors. 

\subsection{Countermeasures against electrical analysis}
\begin{figure}[ht!]
\centering
\includegraphics[width=130mm]{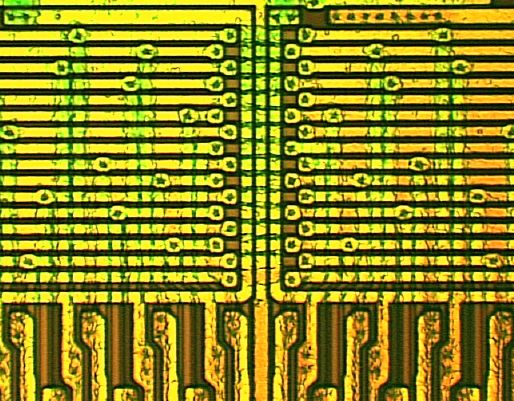}
\caption{Optical microscopic image of a bus scrambler. Although adding some obscurity, the permutation can easily be identified. Source:\cite{doc:kommerling}}
\label{fig:buspermutation}
\end{figure}

A countermeasure that hinders both optical and electrical analysis is obfuscating busses. The busses are the connections between the different 'building blocks' of the chip: memory, control logic, crypto and other functionality is connected this way. Tapping the electrical bus signals is interesting for it allows a high-level view of the communication between different modules on the chip. Obfuscating the bus signals makes it harder to understand the communication between modules. Figure \ref{fig:buspermutation} shows a scrambling block, that 'randomizes' the order of the parallel bus circuits. The additional complexity, however, is easily defeated by visually inspecting the permutation block. 

More advanced solutions are possible, though. One might implement logic that dynamically encrypts or encodes bus signals with continuously changing keys. However, because of the high amounts of extra circuitry required for such solutions, this is a very expensive solution and only appropriate for protecting high-value hardware.

\section{Conclusion}
\label{sec:conclusion}
This paper has established that while RFID technology is increasingly complex, a wide variety of approaches to reverse engineering exist. While the most effective approaches will inevitably destroy the tag in the process, RFID is a technology that is often deployed in a way that it is relatively easy for an attacker to obtain one or more samples. Currently, many RFID solutions are not armoured against reverse-engineering, which is probably both a cost concern as a lack of attention for the potential problems that may arise.

However, reverse-engineering also points out another problem: manufacturers build and deploy RFID solutions without releasing details about its protocol, the internal algorithms and implementation details. This leads to a dangerous situation where a malicious opponent with sufficient skill and means may discover and exploit vulnerabilities in cards of which millions are being used for various purposes. If manufacturers would adopt a more transparent approach to developing and selling products, this problem would at least be more manageable, as vulnerabilities would generally be discovered earlier and, more importantly, would be public knowledge. This way, all potentially interested parties can make an informed choice for an adequate RFID solution. 

\bibliographystyle{splncs}

\begin{thebibliography}{1}

\bibitem{web:merweb-def}
Definition of 'reverse engineering' according to Merriam-Webster dictionary, 
\url{http://www.merriam-webster.com/dictionary/reverse\%20engineering}

\bibitem{web:history}
Wisegeek page on reverse engineering,
\url{http://www.wisegeek.com/what-is-reverse-engineering.htm}

\bibitem{doc:stateoftheartIC}
The State-of-the-Art in IC Reverse Engineering, by Randy Torrance and Dick James,
\url{http://www.iacr.org/archive/ches2009/57470361/57470361.pdf}

\bibitem{web:altertech}
Reverse Engineering \& Construction Analysis, a page on the website of Alter technology, 
\url{http://www.altertechnology.com/atn/en/engineering/reverse-engineering-451.htm}

\bibitem{ebook:sipp}
Software and Intellectual Property Protection: Copyright and Patent Issues, by Bernard A. Galler, 
\url{http://books.google.nl/books?id=QACY2JCu4BUC&pg=PA130&redir_esc=y#v=onepage&q&f=false}

\bibitem{doc:jtag}
Blackbox JTAG Reverse Engineering, by Felix Domke,
\url{http://events.ccc.de/congress/2009/Fahrplan/attachments/1435_JTAG.pdf}

\bibitem{web:eetimes}
The RF in RFID, by Daniel M. Dobkin, 
\url{http://store.elsevier.com/product.jsp?isbn=9780123945839&pagename=search}. A relevant excerpt may be found on 
\url{http://www.eetimes.com/document.asp?doc_id=1276306}

\bibitem{doc:iso14443}
ISO/IEC 14443 Identification cards - Contactless integrated circuit cards - Proximity cards,
\url{http://www.iso.org/iso/home/search.htm?qt=ISO\%2FIEC+14443&sort=rel&type=simple&published=on}. The draft for these specifications are available for free at \url{http://www.waazaa.org/14443/}

\bibitem{web:snifferonly}
Project page of the OpenPICC SnifferOnly, a 13.56MHz RFID sniffer,
\url{http://www.openpcd.org/OpenPICC_SnifferOnly_13.56MHz}

\bibitem{doc:yossen-shamir_poweranalysis}
Power Analysis of RFID Tags, by Yossi Oren and Adi Shamir. Abstract and summary is available on:
\url{http://www.wisdom.weizmann.ac.il/~yossio/rfid/}

\bibitem{doc:hutter_powerEManalysis}
Power and EM Attacks on Passive 13.56MHz RFID Devices, by Michael Hutter, Stefan Mangard and Martin Feldhofer. 
\url{https://www.iacr.org/archive/ches2007/47270320/47270320.pdf}

\bibitem{web:14nm}
Article on preparations for 14nm process by Samsung and GlobalFoundries,
\url{http://www.extremetech.com/computing/181136-samsung-and-globalfoundries-buddy-up-for-14nm-while-ibm-heads-for-the-exit}

\bibitem{doc:kommerling}
Design Principles for Tamper-Resistant Smartcard Processors, by Oliver Kömmerling, 
\url{https://www.usenix.org/legacy/events/smartcard99/full_papers/kommerling/kommerling_html/}

\bibitem{doc:karstennohl}
Reverse-Engineering a Cryptographic RFID Tag, by Karsten Nohl et al., 
\url{http://luci.ics.uci.edu/websiteContent/weAreLuci/biographies/faculty/djp3/LocalCopy/usenix08.pdf}

\bibitem{web:embedded}
Article "Radio Basics for UHF RFID" on embedded.com,
\url{http://www.embedded.com/design/industrial-control/4019132/Tutorial-Radio-Basics-for-UHF-RFID-Part-IV}
\end{thebibliography}

All of these links were confirmed to be reachable on 24-04-2014

\end{document}